\documentclass[12pt,reqno]{amsart}
\usepackage{amsmath}
\usepackage {multirow}
\usepackage {lscape}
\usepackage{appendix}
\usepackage[sorting=nyt, url = false, doi = false, eprint = false, isbn = false]{biblatex}
\usepackage{float}

\usepackage {amsfonts}
\usepackage {amsmath}
\usepackage {amsthm}
\usepackage {amssymb}
\usepackage {cleveref}
\usepackage {framed}
\usepackage {amsxtra}
\usepackage {enumerate}
\usepackage {graphicx}
\usepackage{color}
\usepackage{todonotes}
\usepackage{graphicx}

\usepackage[left=0.9in, right=0.9in, top=1.15 in, footskip=0.5in, bottom=1.15 in]{geometry}
\makeatletter

\theoremstyle{definition}

\theoremstyle{plain}

\newcommand{\bQ}{{\bf Q}}
\newcommand{\bX}{{\bf X}}

\newcommand{\bq}{{\bf q}}

\newcommand{\bp}{{\bf p}}
\newcommand{\bu}{{\bf u}}

\newcommand{\bb}{{\bf b}}

\newcommand{\bk}{{\bf k}}

\newcommand{\bx}{{\bf x}}
\newcommand{\by}{{\bf y}}

\newcommand{\dq}{{\frac{d^3 q}{(2\pi)^3}}}

\newcommand{\dk}{{\frac{d^3 k}{(2\pi)^3}}}

\newcommand{\R}{\mathbb{R}}

\newcommand{\wo}{\omega_0}
\newcommand{\dkh}{{d \hat{\bk}}}
\newcommand{\kh}{\hat{\bk}}
\newcommand{\eps}{\varepsilon}
\newcommand{\ls}{l_s}
\newcommand{\We}{W_\eps}

\title{Kinetic Equations for real Scalar Fields Coupled to a Continuum of Atoms}
\author{Joseph Kraisler}
\address{Department of Mathematics and Statistics, Amherst College, Amherst, MA}
\email{jkraisler@amherst.edu}

\author{Joonsoo Lee}
\address{Department of Applied Physics and Applied Mathematics, Columbia University, New York, NY}
\email{jl5062@columbia.edu }
\begin{document}



\maketitle

\begin{abstract}
We consider a model of a scalar field, with dispersion relation $\omega(\bk)$, coupled to a random medium of two level atoms. We investigate the dynamics of states with at most one quanta of excitation in the system. In a high frequency limit, the probability amplitudes are shown to satisfy kinetic equations. Under additional hypotheses on the dispersion relation, we obtain an analytically solvable expression in the diffusion limit.
\end{abstract}

\section{Introduction}\label{sec:intro}
In ~\cite{kraisler22}, the authors considered a fully quantum optical model of a scalar field coupled to a continuum collection of two level atoms. The field, being a scalar model of a free electromagnetic field, satisfied the dispersion relation $\omega(\bk)=c\vert\bk\vert$. The number density of these atoms was assumed to be a random field with small fluctuations around a mean density $\rho_0$. By implementing a formal multiple-scale asymptotic analysis~\cite{Bal2005,Carminati_2020,Caze_2015,Ryzhik_1996}, the authors derived both radiative transport equations (RTE) and diffusion approximations (DA) satisfied by the atomic and field probability densities in a high frequency limit. The photonic model has been studied analytically and numerically in the case of a discrete set of atoms~\cite{Hoskins21,Hoskins23, Mirza_2016}, parallel chiral waveguides~\cite{Hoskins:23}, and for higher numbers of photons ~\cite{kraisler23}.  \\

In this work, we generalize the results of ~\cite{kraisler22} to the case that the scalar field no longer represents a photonic electromagnetic field. To do this we replace the disperison relation $\omega(\bk)=c\vert\bk\vert$ by a general smooth function $\omega(\bk)$. Such a model may then represent any number of quantum particles (or quasiparticles) coupled to a highly disordered distribution of atoms, such as phonon-polaritons, electron-holes, or polarons~\cite{kittel2004introduction}. Specifically, we study the $2\times 2$ system of pseudo-differential equations
\begin{subequations} \label{eq:dynamics}
\begin{align} 
i\partial_t \psi &= \omega(-i \nabla) \psi + g \rho(x) a \, ,\label{eq:dynamics1} \\
i \partial_t a &= g \psi + \Omega a  \, . \label{eq:dynamics2}
\end{align}
\end{subequations}
where the operator $\omega(-i\nabla)$ is defined via the Fourier transform and the dispersion relation $\omega(\bk)$. The atomic number density, $\rho(\bx)$, is assumed to be a random field with small fluctuations about a constant mean density $\rho_0>0$. From this model we obtain a pair of radiative transport equations (RTEs) satisfied by a phase space representation of the amplitudes $\psi(\bx,t)$ and $a(\bx,t)$. Additionally, after placing additional constraints on the dispersion relation, $\omega(\bk)$, we obtain analytically solvable equations for the average probability densities $\langle\vert \psi(\bx,t)\vert^2\rangle$ and $\langle\vert\alpha(\bx,t)\vert^2\rangle$ as $\|\bx\|,\vert t\vert\to\infty$ which we refer to as the diffusion approximation (DA).\\

The rest of the paper is structured as follows. In \Cref{sec:model} we discuss the model in more detail as well as the mathematical tools used to introduce the stochastic features of the system and the phase space representation used to study it. In \Cref{sec:derivation} we use a multiple-scale analysis to derive the RTEs, the main result of the paper. Afterwards, in \Cref{sec:DA} we place additional restrictions on the dispersion relation which allow us to obtain analytically solvable equations in certain spatiotemporal regimes. Finally in \Cref{sec:comparison}, we compare these results to those obtain for the photonic model~\cite{kraisler22}. Many of the detailed calculations have been placed in various appendices. Moreover, we have made a concerted effort not to reproduce calculations which appear in ~\cite{kraisler22} exactly and refer the reader to that paper whenever necessary.

\section{Description of Model and Mathematical Tools} 
\label{sec:model}

We consider the following model of a scalar quantized field coupled to a collection of two level atoms. Let $\omega(\bk)$ be the dispersion relation of a scalar quantized field. We assume that $\omega$ is real valued and smooth away from $\bk=0$. Then we introduce the analog of the real space Hamiltonian from \cite{kraisler22}:
\begin{align} \label{eq:Hamiltonian}
    \mathcal{H} = \int_{\R^3} \hbar\omega(i\nabla)\phi^{\dagger}(\bx)\phi(\bx) + \hbar\Omega\sigma^{\dagger}(\bx)\sigma(\bx)\rho(\bx) + \hbar g(\sigma(\bx)\phi^{\dagger}(\bx)+\sigma^{\dagger}(\bx)\phi(\bx))\rho(\bx) d^3 x.
\end{align}

This Hamiltonian is comprised of three pieces: the total energy in the electromagnetic field, the total energy in the collection of two level atoms, and an interaction energy term which is bilinear in the creation and raising operators. Here the operator $\omega(i\nabla)$ acts on Schwartz functions via the Fourier transform
\begin{align}
    \omega(i\nabla)f(\bx) = \int_{\R^3} e^{i\bk\cdot\bx}\omega(\bk)\tilde{f}(\bk)\dk .
\end{align}
Additionally, the operators $\phi(\bx)$ and $\sigma(\bx)$ satisfy commutation and anticommutation relations respectively
\begin{align}
    [\phi(\bx),\phi^{\dagger}(\by)] &= \delta(\bx-\by)\ , &[\phi(\bx),\phi(\by)] &= 0 , \\
    \{\sigma(\bx),\sigma^{\dagger}(\by)\} &= \frac{\delta(\bx-\by)}{\rho(\bx)}\ , & \{\sigma(\bx),\sigma(\by)\} &= 0 .
\end{align}
We consider the dynamics associated to the Hamiltonian (\ref{eq:Hamiltonian}) with at most one excitation present in the system. The equations of motion, derived in a similar manner as those in Appendix A of \cite{kraisler22}, are
\begin{align}
i\partial_t \psi &= \omega(-i \nabla) \psi + g \rho(x) a \, , \\
i \partial_t a &= g \psi + \Omega a  \, .
\end{align}

\subsection{Random Medium} \label{subsec:randomness}

In this section we describe the model of randomness that we implement. Assume the atomic density $\rho(\bx)$ is of the form
\begin{align} \label{eq:randomrho}
    \rho(\bx) = \rho_0(1+\eta(\bx))\, , 
\end{align}
where $\rho_0$ is constant and $\eta(\bx)$ is a mean zero real valued random field with correlation function $C(x)$. More precisely, defining $\langle \cdots\rangle = \mathbb{E}[\cdots]$ to be the expectation over realizations of the medium,
\begin{align}
    \langle\eta(\bx)\rangle & = 0 \, , \\
    \langle\eta(\bx)\eta(\by)\rangle & = C(\vert\bx-\by\vert)\, .
\end{align}
Notice that we have chosen an isotropic and statistically homogeneous medium where the correlations depend only on the distance between two points.

\subsection{Wigner Transform} \label{subsec:Wigner}

One of the main tools we use in this work is the \textit{scaled Wigner transform}, which allows us to obtain a local conservation law for the intensity of the field $\bu$ that is resolved over both position and direction. For each fixed function $\bu = (u_1,u_2)$, The Wigner transform, $W_{\eps}(\bx, \bk, t)$, is defined:
\begin{align} \label{eq:scaledWigner}
    W_{\eps}(\bx, \bk, t) = \int \frac{d^3 x'}{(2 \pi)^3} e^{-i \bk \cdot\bx'} \bu (\bx - \eps\bx'/2 , t) \bu^\dagger (\bx + \eps\bx'/2 , t)\, .
\end{align}
The Wigner transform has several useful properties. It is real valued and the probability densities $|u_1(\bx, t)|^2$ and $|u_2(\bx, t)|^2$ can be recovered from the Wigner transform by the relations
\begin{align} \label{eq:prob1}
    |u_1(\bx, t)|^2 &= \int \dk (W_{\eps})_{11}(\bx, \bk, t), \\
    |u_2(\bx, t)|^2 &= \int \dk (W_{\eps})_{22}(\bx, \bk, t). \label{eq:prob2}
\end{align}
Note that in the above expressions $W_{ij}(\bx,\bk,t)$ is the $i$,$j$ component of the $2\times 2$ Wigner transform $W(\bx,\bk,t)$.

\section{Derivation of Kinetic Equations} \label{sec:derivation}

In this section, starting at the system (\ref{eq:dynamics}), we derive a radiative transport equation (RTE) satisfied by the components of the Wigner transform in a certain basis. From this RTE, it is possible to obtain the probability densities through the relations \eqref{eq:prob1} and \eqref{eq:prob2}. 

\subsection{Liouville Equation} \label{subsec:Liouville}
We begin by defining the vector $\bu(\bx, t) = 
\begin{bmatrix}
    \psi(\bx, t) ,
    a(\bx, t) \sqrt{\rho_0}
\end{bmatrix}^\top$ 
and recall our choice of random field $\rho(\bx) = \rho_0 (1 + \eta(\bx))$. From the original system \ref{eq:dynamics} we find that $\bu$ satisfies
\begin{align} \label{eq:dynamics3}
    i \partial_t \bu = A(\bx) \bu + g \sqrt{\rho_0} \eta(\bx) K \bu\, ,
\end{align}
where the operators $A(\bx)$ and $K$ are defined
\begin{align}
    A(\bx) &= \begin{bmatrix}
     \omega(-i \nabla_\bx) & g\sqrt{\rho_0} \\
        g\sqrt{\rho_0} & \Omega
    \end{bmatrix}\, , \\
    K &= \begin{bmatrix}
        0 & 1 \\ 0 & 0
    \end{bmatrix}.
\end{align}

Next we introduce a small parameter $\eps$ and rescale the temporal and spatial variables as $t \rightarrow t/\eps$, $\bx \rightarrow \bx/\eps$ . In addition, we assume that the randomness is sufficiently weak so that the correlation function $C$ is $O(\eps)$. To this end, we rescale $\eta\to\sqrt{\eps}\eta$. Note that $\nabla_{\bx/\eps} = \eps \nabla_\bx$, which gives us the equation
\begin{gather}
    \eps i \partial_t \bu_\eps = A_\eps (\bx) \bu_\eps + \sqrt{\eps} g \sqrt{\rho_0} \eta(\bx/\eps) K \bu_\eps, \\
    A_\eps = \begin{bmatrix}
         \omega(-\eps i \nabla_\bx) & g\sqrt{\rho_0} \\
        g\sqrt{\rho_0} & \Omega
    \end{bmatrix}.
\end{gather}
By following similar steps as in Appendix D in \cite{kraisler22}, it can be shown that the scaled Wigner transform, $\We(\bx,\bk,t)$ , given in \Cref{eq:scaledWigner} satisfies the following Liouville equation: 
\begin{multline} \label{eq:Liouville1}
    \eps i \partial_t \We(\bx, \bk, t) = \int \frac{d^3q}{(2\pi)^3}e^{i \bq \cdot \bx} [\Tilde{A}_\eps(-\bk/\eps + \bq/2) \Tilde{W}_\eps(\bq, \bk, t) - \Tilde{W}_\eps(\bq, \bk, t) \Tilde{A}_\eps(-\bk/\eps - \bq/2)] \\ +
    \sqrt{\eps} g \sqrt{\rho_0} \int \frac{d^3 q}{(2 \pi)^3} e^{i \bq \cdot \bx/\eps} \Tilde{\eta}(\bq) [K \We (\bx, \bk + \bq/2, t) - \We(\bx, \bk - \bq/2, t) K^T]\, .
\end{multline}
The operator $\tilde{A}_{\eps}$ is the Fourier transform of the operator valued matrix $A_{\eps}$ and is given by
\begin{align*}
    \Tilde{A}_\eps(\bk) = \begin{bmatrix}
        \omega (\eps \bk) & g \sqrt{\rho_0} \\
        g \sqrt{\rho_0} & \Omega
    \end{bmatrix}\, .
\end{align*}
\subsection{Multiple Scale Asymptotics} \label{subsec:MSA}

Now consider the behavior of $\We$ under a high-frequency limit $\eps \rightarrow 0$, which allows us to separate the problem into microscopic and macroscopic scales. We first define a fast variable and then examine the different orders of the asymptotic expression for $\We$, resulting in a system of equations. The solution of this system is will give us a general RTE. First, defining the fast variable
\begin{align}
\bX = \frac{\bx}{\eps}.
\end{align}
Treating $\bX$ and $\bx$ as independent variables, the chain rule results in the following transformation:
\begin{align}
    \nabla_\bx \mapsto \nabla_\bx + \frac{1}{\eps} \nabla_\bX.
\end{align}
By assuming the Wigner transform $\We$ depends on both the fast and slow variables, and performing the above substitution, the Liouville equation \eqref{eq:Liouville1} becomes
\begin{multline} \label{eq:Liouville2}
    \eps i \partial_t \We(\bx, \bX, \bk, t) = \int \frac{d^3 q}{(2\pi)^3} \frac{d^3 Q}{(2\pi)^3} e^{i\bq\cdot\bx + i\bQ\cdot\bX}[\Tilde{A}_\eps(-\bk/\eps+\bq/2+\bQ/2\eps)\Tilde{W}_\eps(\bq, \bQ, \bk, t)\\ -
    \Tilde{W}_\eps(\bq, \bQ, \bk, t)\Tilde{A}_\eps(-\bk/\eps-\bq/2-\bQ/2\eps)] \\ +
    \sqrt{\eps} g \sqrt{\rho_0} \int \frac{d^3 q}{(2 \pi)^3} e^{i \bq \cdot \bX} \Tilde{\eta}(\bq) [K \We (\bx, \bX, \bk + \bq/2, t) - \We(\bx, \bX, \bk - \bq/2, t) K^T].
\end{multline}
Next, assume that the Wigner transform $\We$ has an asymptotic expansion in powers of $\sqrt{\eps}$,
\begin{align}\label{eq:Wignerexpansion}
    \We(\bx, \bk, t) = W_0(\bx, \bk, t) + \sqrt{\eps}W_1(\bx, \bX, \bk, t) + \eps W_2(\bx, \bX, \bk, t) + ...,
\end{align}
where the first term $W_0$ is assumed to be deterministic and independent of the fast variable $X$. Next, we expand the dispersion relation $\omega(\bk)$ via Taylor's theorem:
\begin{align}\label{eq:dispersionexpansion}
    \omega(-\bk + \eps\bq/2 + \bQ/2) = \omega(-\bk + \bQ/2) + \nabla \omega(-\bk + \bQ/2) \cdot \eps\bq/2 + O(\eps^2).
 \end{align}
We substitute \Cref{eq:dispersionexpansion} and \Cref{eq:Wignerexpansion} into the Liouville Equation \ref{eq:Liouville2} derived earlier. By equating coefficients of each power of $\eps$, we obtain a hierarchy of equations:
\begin{align}\label{eq:o1}
   O(1): \Tilde{A}_\eps(-\bk/\eps)W_0(\bx, \bk, t) - W_0(\bx, \bk, t)\Tilde{A}_\eps(-\bk/\eps) = 0.
\end{align}
\begin{align}\label{eq:osqrteps}
\begin{split}
    O(\sqrt{\eps}): \Tilde{A}_\eps (-(\bk - \bQ/2)/\eps) \Tilde{W}_1(\bx, \bQ, \bk, t) - \Tilde{W}_1(\bx, \bQ, \bk, t) \Tilde{A}_\eps (-(\bk + \bQ/2)/\eps) \\= 
    g \sqrt{\rho_0} \Tilde{\eta}(\bq) [W_0(\bx, \bk - \bQ/2, t)K^T - K W_0(\bx, \bk + \bQ/2, t)].
\end{split}
\end{align}
\begin{align} \label{eq:oeps}
\begin{split} 
   O(\eps): i\partial_tW_0 = LW_2 + g\sqrt{\rho_0} \int \frac{d^3 q}{(2\pi)^3} e^{i\bq\cdot\bX} \Tilde{\eta}(\bq) [K W_1(\bx, \bx, \bk + \bq/2, t) - W_1(\bx, \bx, \bk - \bq/2, t)K^T] \\
    + \int \frac{d^3 q}{(2\pi)^3} \frac{d^3 Q}{(2\pi)^3} e^{i\bq\cdot\bx + i\bQ\cdot\bX} [\Tilde{M}(\bq/2, -\bk + \bQ / 2) \tilde{W}_0(\bq, \bk, t) + \Tilde{W}_0(\bq, \bk, t) \Tilde{M}(-\bq/2, -\bk - \bQ / 2)],
\end{split}
\end{align}
where 
\begin{align}
    L\We(\bx, \bk) &= \int \dq e^{i\bq \cdot \bx/\eps} \tilde{\eta}(\bq) [\We(\bx, \bk + \bq/2) - \We(\bx, \bk - \bq/2)]\ , \\
    \Tilde{M}(\bq/2, -\bk + \bQ / 2) &= \begin{bmatrix}
        \nabla \omega(-\bk + \bQ/2)\cdot \bq/2& 0 \\
        0 & 0
    \end{bmatrix}
\end{align}
\Cref{eq:o1} says that $W_0$ and $\Tilde{A}_{\eps}$ are simultaneously diagonalizable. We can then expand $W_0$ as 
\begin{equation} \label{eq:W0expansion}
    W_0(\bx, \bk, t) = a_+(\bx, \bk, t)\bb_+(\bk)\bb_+^T(\bk) + a_-(\bx, \bk, t)\bb_-(\bk)\bb_-^T(\bk),
\end{equation}
where we define:
\begin{align}
    \lambda_\pm(\bk) = \frac{(c \omega(-\bk) + \Omega) \pm \sqrt{(c \omega(-\bk) - \Omega)^2 + 4g^2\rho_0}}{2},
\end{align}
\begin{align}
    \bb_\pm(\bk) = \frac{1}{\sqrt{(\lambda_\pm - \Omega)^2 + g^2\rho_0}}\begin{bmatrix}
        \lambda_\pm - \Omega \\
        g\sqrt{\rho_0}
    \end{bmatrix}, 
\end{align}
The next two equations are more substantial and we show in Appendix \ref{appx:b}, that the coefficients $a_{\pm}$ in the expansion \Cref{eq:W0expansion} satisfy the following kinetic equations.
\begin{multline} \label{eq:Kinetic1}
     \frac{1}{c}\partial_t a_\pm (\bx, \bk, t) +  f_\pm (\bk) \nabla \omega(-\bk) \cdot \nabla_\bx a_\pm(\bx, \bk, t) = \\
     \zeta_\pm(\bk) \int \frac{d^3q}{(2\pi)^3}\delta(\omega(-\bq) - \omega(-\bk)) \Tilde{C}(\bk - \bq) [a_\pm(\bx, \bk, t) - a_\pm(\bx, \bq, t)],
\end{multline}
\begin{align}
    f_\pm(\bk) &= \frac{(\lambda_\pm(\bk) - \Omega)^2}{(\lambda_{\pm}(\bk) - \Omega)^2 + g^2\rho_0}, \\
    \zeta_\pm(\bk) &= \frac{2\pi (g^2 \rho_0 )^2|\lambda_\pm (\bk) - \Omega|}{((\lambda_\pm (\bk) - \Omega)^2 + g^2\rho_0)^2} \sqrt{( \omega(-\bk) - \Omega)^2 + 4 g^2 \rho_0}.
\end{align}
This equation is the main result of the paper for a general smooth$\omega(\bk)$. We obtain the Wigner transform $W_0$ from the RTE above and then find the average probability densities $\langle|\psi|^2\rangle$ and $\langle|a|^2\rangle$,
\begin{align} \label{eq:AvgPsi}
    \langle|\psi(\bx,t)|^2\rangle  &= \int \dk \left(\frac{a_+(\bx, \bk, t)(\lambda_+(\bk)-\Omega)^2}{(\lambda_+(\bk)-\Omega)^2 + g^2 \rho_0} + \frac{a_-(\bx, \bk, t)(\lambda_-(\bk)-\Omega)^2}{(\lambda_-(\bk)-\Omega)^2 + g^2 \rho_0}\right), \\\label{eq:AvgA}
     \langle|a(\bx,t)|^2\rangle  &= \int \dk \left(\frac{a_+(\bx, \bk, t)}{(\lambda_+(\bk)-\Omega)^2 + g^2 \rho_0} + \frac{a_-(\bx, \bk, t)}{(\lambda_-(\bk)-\Omega)^2 + g^2 \rho_0}\right).
\end{align}

\section{Diffusion approximation} \label{sec:DA}
In order to arrive at a theory of diffusion in a reasonable spatial and temporal limit, we make the following two assumptions about the dispersion relation $\omega(\bk)$:
\begin{enumerate}
    \item $\omega(\bk)$ only depends on the magnitude of $k$, $|\bk|$. Equivalently, there exists a function $\wo:[0,\infty)\to\R$ such that
        \begin{align}
            \omega(\bk) = \wo(|\bk|).
        \end{align}
    \item The function $\wo:(0,\infty)\to\R$ is smooth with nonvanishing derivative
\begin{align}
    \wo'(k) \neq 0\, .
\end{align}
\end{enumerate}
By using the identity
\begin{align}
    \delta(\wo(|\bq|) - \wo(|\bk|)) = \frac{\delta(|\bq| - |\bk|)}{|\wo'(|\bk|)|},
\end{align}
along with the definitions
\begin{align}
    F(\bk) &= f(\bk) \wo'(|\bk|), \\
    \mu(\bk) &= \frac{\zeta(\bk)}{|\wo'(|\bk|)|}|\bk|^2 \int d\hat{k}' \tilde{C}(|\bk|(\hat{\bf{k}} - \hat{\bf{k}}')), \\
    A(\bk, \bk') &= \frac{\Tilde{C}(|\bk|(\hat{\bk} - \hat{\bk}'))}{\int d\hat{\bf{k}}' \Tilde{C}(|\bk|(\hat{\bk} - \hat{\bk}'))},
\end{align}
we obtain the following RTE from \Cref{eq:Kinetic1}:
\begin{align} \label{eq:RTE}
\begin{split}
     \frac{1}{c}\partial_t a_\pm (\bx, \bk, t) &+ F_{\pm}(\bk) \hat{\bk} \cdot \nabla_\bx a_\pm(\bx, \bk, t)\\
     &+ \mu(\bk) a_{\pm}(\bx, \bk, t) = 
     \mu(\bk) \int \dkh' A(\bk, \bk') a_{\pm}(\bx, \bk', t).
\end{split}
\end{align}

As in \cite{kraisler22}, we compute a diffusion approximation for the RTE from \Cref{eq:RTE}. First, note that the RTE \Cref{eq:RTE} takes the form:
\begin{align}
    \frac{1}{c}\partial_t I(\bx, \bk, t) + F_{\pm}(\bk) \hat{\bk} \cdot \nabla_\bx I(\bx, \bk, t) + \mu_s I(\bx, \bk, t) = \mu_s LI(\bx, \bk, t).
\end{align}
From \cite{Carminati_2020}, a diffusion approximation is obtained by expanding $I$ in spherical harmonics. To lowest order, it can be shown that
\begin{align}
    I(\bx, \bk, t) = \frac{c}{4\pi}\left( u - \ell^* \hat{\bk} \cdot \nabla u \right).
\end{align}
where we define
\begin{gather} \label{eq:angularintegral}
    u(\bx, |\bk|, t) = \int d\hat{\bk}I(\bx, \bk, t), \\
    \ell^* = \frac{F_\pm (\bk)^2}{(1-g)\mu_\pm(\bk)}.
\end{gather}
Then, $u(\bx,\vert\bk\vert,t)$ satisfies the diffusion equation
\begin{gather}
    \partial_t u - D\Delta u = 0, \\
    D = \frac{1}{3}c \ell^*.
\end{gather}
As the solution of $\partial_t u - D \Delta u=0$ in an infinite medium is given by
\begin{align}
    u(\bx, t) = \frac{1}{(4\pi D t)^{\frac{3}{2}}} \int d^3 \bx' \exp\left(\frac{-|\bx - \bx'|^2}{4Dt}\right) u(\bx', 0),
\end{align}
we can solve this equation exactly for given initial conditions. Then, using the relationships, \Cref{eq:prob1},  \Cref{eq:angularintegral}, and \Cref{eq:W0expansion}, we obtain equations for the average probability densities $\langle\vert\psi(\bx,t)\vert^2\rangle$ and $\langle\vert a(\bx,t)\vert^2\rangle$:
\begin{align}
    \rho_0 \langle|\psi(\bx,t)|^2\rangle  &= \int_{0}^{\infty} dk k^2 \left( \frac{u_+(\bx, k, t)(\lambda_+(k)-\Omega)^2}{(\lambda_+(k)-\Omega)^2 + g^2 \rho_0} + \frac{u_-(\bx, k, t)(\lambda_-(k)-\Omega)^2}{(\lambda_-(k)-\Omega)^2 + g^2 \rho_0} \right), \\
     \langle|a(\bx,t)|^2\rangle  &= \int_0^{\infty} dk k^2 \left( \frac{u_+(\bx, k, t)}{(\lambda_+(k)-\Omega)^2 + g^2 \rho_0} + \frac{u_-(\bx, k, t)}{(\lambda_-(k)-\Omega)^2 + g^2 \rho_0} \right).
\end{align}

\section{Case Study: The Phonon-Polariton}\label{sec:comparison}
In this section, we consider a particular example of the situation described in Section (\ref{sec:model}). Phonon-Polaritons are quasiparticles which arise from the coupling between optical phonons and infrared photons. In some wavenumber regimes the dispersion relation of these quasiparticles can be modeled as a quadratic function of $\vert\bk\vert$. To simplify the results we assume that this is valid for every mode. Specifically, suppose
\begin{align}
    \omega(\bk) = c_0\vert\bk\vert^2 ,
\end{align}
for some $c_0\in \R$. This means that our starting point, system (\ref{eq:dynamics1}), becomes
\begin{align}
    i \partial_t \psi &= -c_0 \Delta \psi + g \rho(\bx) a, \\ 
    i \partial_t a &= g \psi + \Omega a.
\end{align}
Suppose that initially, the energy is entirely in a collection of excited atoms near the origin. One possible formula for such an initial set up is given by 
\begin{align}\label{eq:icpa}
a(\bx, 0) &= \left(\frac{1}{\pi l_s}\right)^{3/4} e^{-|\bx|^2/2l_s^2}, \\
\psi(\bx, 0) &= 0. 
\end{align}
Within the diffusion approximation discussed in \Cref{sec:DA}, the functions $u_{\pm}(\bx,\vert\bk\vert,t)$ satisfy
\begin{align}
     \partial_t u_{\pm} - D_{\pm} \Delta u_{\pm} = 0, \label{eq:da} \\
    D_{\pm} = \frac{c F_\pm (\bk)^2}{3(1-g)\mu_\pm(\bk)}.
\end{align}
The functions which comprise the diffusion coefficients $D_{\pm}$ are given by
\begin{align}
     \lambda_\pm(\bk) &= \frac{(c_0 |\bk|^2 + \Omega) \pm \sqrt{(c |\bk|^2 - \Omega)^2 + 4g^2 \rho_0}}{2}, \\
    F(\bk) &= \frac{2(\lambda(\bk) - \Omega)^2|\bk|}{(\lambda(\bk) - \Omega)^2 + g^2 \rho_0}, \\
    \mu_\pm(\bk) &= \frac{2\pi(g^2 \rho_0)^2 |\lambda_\pm(\bk) - \Omega|}{c_0^2 ((\lambda_\pm(\bq) - \Omega)^2 + g^2 \rho_0)^2}\sqrt{(c_0|\bk|^2 - \Omega)^2 + 4g^2\rho_0} \frac{|\bk|}{2}\int \frac{\dkh'}{(2\pi)^3} \Tilde{C}(|\bk|(\kh - \kh')).
\end{align}
The initial conditions for $\psi$ and $a$ given in \Cref{eq:icpa} impose conditions on $u_{\pm}$ 
\begin{align}\label{eq:icu}
    u_\pm(\bx, |\bk|, 0) = \frac{4}{\pi^2} \frac{(\lambda_\pm - \Omega)^2 + g^2 \rho_0}{(\lambda_\mp(\bk) - \Omega)^2 - (\lambda_\pm(\bk) - \Omega)^2} \frac{(\lambda_\pm - \Omega)^2}{g^2 \rho_0}e^{-l_s^2 |\bk|^2}e^{-|\bx|^2/l_s^2}.
\end{align}
Solving the diffusion equation \Cref{eq:da} with these initial conditions leads to.
\begin{multline}
        u_\pm(\bx, |\bk|, t) = \frac{4}{\pi^2} \frac{(\lambda_\pm(\bk) - \Omega)^2 + g^2 \rho_0}{(\lambda_\mp(\bk) - \Omega)^2 - (\lambda_\pm(\bk) - \Omega)^2} \frac{(\lambda_\pm(\bk) - \Omega)^2}{g^2 \rho_0}e^{-l_s^2 |\bk|^2} \\
        \times \left(\frac{l_s^2}{l_s^2 + 4 t D_{\pm}(\bk)}\right)^{3/2} e^{-|\bx|^2/(l_s^2 + 4tD_{\pm}(\bk))}.
\end{multline}
Finally, after integrating over the radial variable $\vert\bk\vert$, the average probability densities are given by
\begin{align} \label{eq:k^2psi}
\begin{split}
    &\langle |\psi(\bx, t)|^2 \rangle \\
    &= \frac{4}{g^2 \rho_0 \pi^2} \int_0^\infty dk k^2 e^{-\ls^2 k^2}\left[\frac{[(\lambda_-(k) - \Omega)(\lambda_+(k) - \Omega)]^2}{(\lambda_-(k) - \Omega)^2 - (\lambda_+(k) - \Omega)^2}\left(\frac{\ls^2}{\ls^2 + 4tD_+(k)}\right)^{3/2} e^{-|\bx|^2/(l_s^2 + 4tD_+(\bk))} \right.\\
    &\left.+ \frac{[(\lambda_+(k) - \Omega)(\lambda_-(k) - \Omega)]^2}{(\lambda_+(k) - \Omega)^2 - (\lambda_-(k) - \Omega)^2}\left(\frac{\ls^2}{\ls^2 + 4tD_-(k)}\right)^{3/2} e^{-|\bx|^2/(l_s^2 + 4tD_-(\bk))}\right]. 
\end{split}
\end{align}
\begin{align} \label{eq:k^2a}
\begin{split}
    &\rho_0 \langle |a(\bx, t)|^2 \rangle\\
    &= \frac{4}{\pi^2} \int_0^\infty dk k^2 e^{-\ls^2 k^2} \left[ \frac{(\lambda_-(k) - \Omega)^2}{(\lambda_-(k) - \Omega)^2 - (\lambda_+(k) - \Omega)^2}\left(\frac{\ls^2}{\ls^2 + 4tD_+(k)}\right)^{3/2} e^{-|\bx|^2/(l_s^2 + 4tD_+(\bk))} \right.\\
    &\left.+ \frac{(\lambda_+(k) - \Omega)^2}{(\lambda_+(k) - \Omega)^2 - (\lambda_-(k) - \Omega)^2}\left(\frac{\ls^2}{\ls^2 + 4tD_-(k)}\right)^{3/2} e^{-|\bx|^2/(l_s^2 + 4tD_-(\bk))} \right]. 
\end{split}
\end{align}
\subsection{Numerical Results}
\begin{figure}[H]
\centering
\includegraphics[scale=0.7]{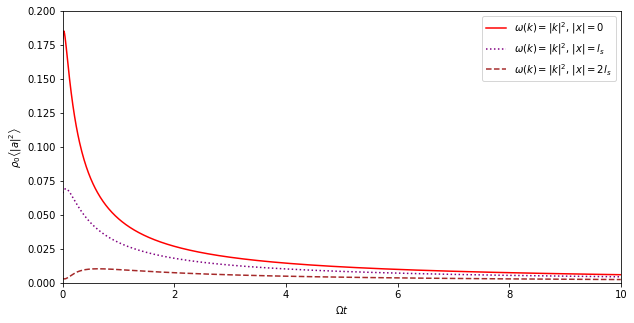}
\caption{Time dependence of atomic probability density in a random medium for several distances.}
\label{fig:k^2}
\end{figure}

\begin{figure}[H]
\centering
\includegraphics[scale=0.7]{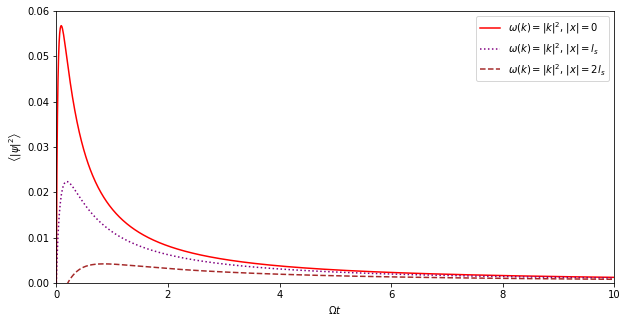}
\caption{Time dependence of field probability density in a random medium for several distances
}
\label{fig:PsiChart}
\end{figure}

Here we provide some plots of the amplitudes given in \Cref{eq:k^2psi} and \Cref{eq:k^2a}. We consider the case of a quadratic dispersion relation $\omega(k) = |\bk|^2$, in an istropic scattering medium for which the scattering amplitude $A(\bk,\bk')$ is constant: $A = 1/(4\pi)$. Note that \cref{eq:k^2psi} and \cref{eq:k^2a} have the same asymptotic behavior with respect to $|\bx|$ and $t$, simply with different constant factors. Additionally, we nondimensionalize the problem,setting the dimensionless quantities $\Omega l_s^2/c = \rho_0 g^2 / \Omega^2 = 1$. Figures \ref{fig:k^2} and \ref{fig:PsiChart} are plots for our implementation of the approximate formula given in \cref{eq:k^2psi} and \cref{eq:k^2a}.
As expected, at larger distances away from the initial volume of excitation, the initial probability is larger and the values decay quicker. It is important to note that the figures may not be accurate for small times due to the breakdown of the diffusion approximation, which can be seen through the function increasing for small t around the origin, as well as the function becoming negative in Figure \ref{fig:PsiChart}.


\subsection{Comparison with Photonic Case}
\begin{figure}[H]
\centering
\includegraphics[scale=0.7]{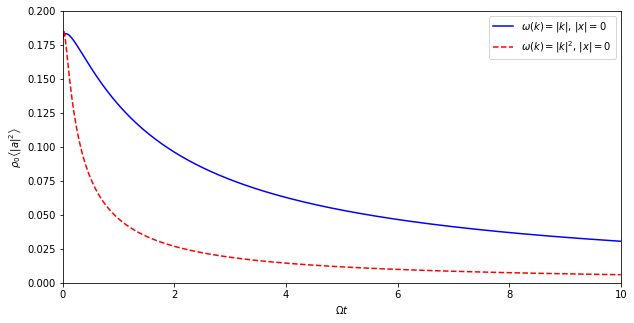}
\caption{Comparison of atomic probability density of linear and quadratic $\omega$ at $|x| = 0$.}
\label{fig:x=0}
\end{figure}

\begin{figure}[H]
\centering
\includegraphics[scale=0.7]{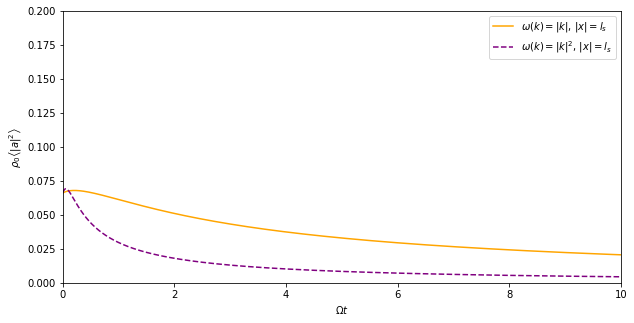}
\caption{Comparison of atomic probability density of linear and quadratic $\omega$ at $|x| = l_s$.}
\label{fig:x=l}
\end{figure}

\begin{figure}[H]
\centering
\includegraphics[scale=0.7]{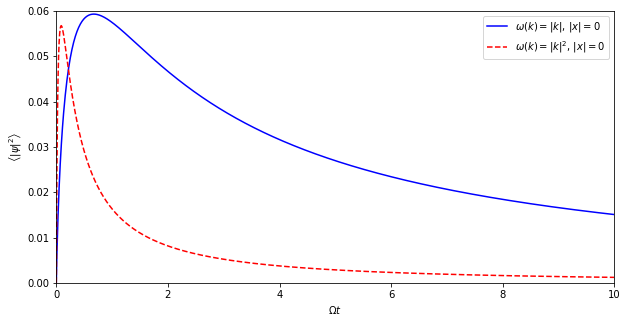}
\caption{Comparison of field probability density of linear and quadratic $\omega$ at $|x| = 0$.}
\label{fig:x=psi0}
\end{figure}

\begin{figure}[H]
\centering
\includegraphics[scale=0.7]{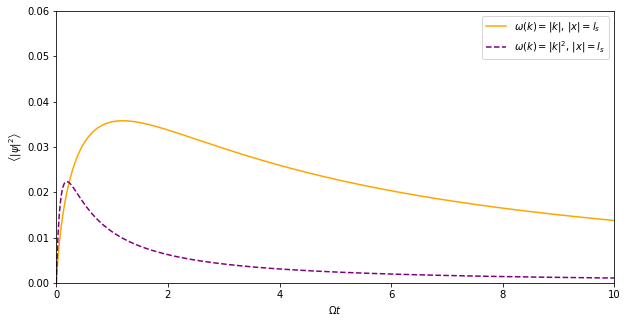}
\caption{Comparison of field probability density of linear and quadratic $\omega$ at $|x| = l_s$.}
\label{fig:x=psil}
\end{figure}

In this section we compare the results for the polariton model with those of the photonic model studied in ~\cite{kraisler22}. We will look at a comparison of the probability densities for the two cases at several distances from the origin. For this Photonic case, we have $\omega(k) = |\bk|$, with $A = 1/(4\pi)$ and nondimensionalize the problem. Then we put the dimensionless quantities $\Omega l_s/c = \rho_0 g^2 / \Omega^2 = 1$. Note that the photonic and polaritonic amplitudes satisfy similar equations with respect to $\bx$ and $t$, just with different constants. Therefore, they have the same asymptotic rate of decay. The Figures \ref{fig:x=0} and \ref{fig:x=l} show comparisons between the atomic probability density of the quadratic case $\omega(k) = |\bk|^2$ and the linear case $\omega(k) = |\bk|$ at different distances from the origin. In Figure \ref{fig:x=0}, we look at the behavior at the origin. In Figure \ref{fig:x=l}, we look at the behavior at displacement $l_s$ from the origin. At both these distances, the linear dispersion relation results in a slower decay compared to the quadratic relation. Figures \ref{fig:x=psi0} and \ref{fig:x=psil} show comparisons between the field probability density of the quadratic case $\omega(k) = |\bk|^2$ and the linear case $\omega(k) = |\bk|$ at different distances from the origin.


\section{Discussion} \label{sec:discussion}

In this paper we considered a scalar quantized Bosonic field of smooth dispersion relation $\omega(\bk)$ which is coupled to a collection of two level atoms. The number density of these atoms is assumed to be a random field with mean $\rho_0>0$ and with correlations on the order of a small parameter $\eps$. In the high frequency limit $\eps\to 0$, we obtain Radiative Transport Equations (RTEs) which describe the probability density functions associated to single excitation modes of the field and atomic states. With some additional hypotheses on the dispersion relation $\omega(\bk)$, we derive analytically solvable equations in a Diffusion Approximation (DA). We use this general framework to study the problem of polaritons with quadratic dispersion relation $\omega(\bk)=c_0\vert\bk\vert^2$ and compare the rate of collective emission to that observed in the photonic case of $\omega(\bk)=c\vert\bk\vert$.

\subsection*{Acknowledgements} This research was supported in part by Simons Foundation Math + X Investigator Award \#376319 (JK).
\appendix

\section{Real space Hamiltonian Eq.~(\ref{eq:Hamiltonian}) and System (\ref{eq:dynamics1})}

The Hamiltonian, \eqref{eq:Hamiltonian}, is obtained in a similar manner to Eq.~(19) in ~\cite{kraisler22}. In fact, the second and third terms are identical. Only the first time, comprising the energy in the scalar field must be altered. We outline the steps here. A scalar quantized field with dispersion relation $\omega(\bk)$ and creation and annhilation operators $a_{\bk}^{\dagger}$ and $a_{\bk}$ satisfying commutation relations
\begin{align}
    [a_{\bk},a_{\bk'}^{\dagger}] = \delta(\bk-\bk'),\quad [a_{\bk},a_{\bk}] = 0\ ,
\end{align}
has an associated Hamiltonian
\begin{align}\label{eq:DickeHamiltonian}
    H_F = \int_{\R^3}\hbar\omega(\bk)a_{\bk}^{\dagger}a_{\bk}\frac{d\bk}{(2\pi)^3}\ .
\end{align}
If we introduce the real space representation
\begin{align}
    \phi(\bx) = \int_{\R^3} e^{i\bx\cdot\bk}a_{\bk}\frac{d\bk}{(2\pi)^{3/2}}\ ,
\end{align}
then $H_F$ can be recast in terms of the variable $\bx$ as
\begin{align}
    H_F = \int_{\R^d}\hbar \omega(i\nabla)\phi^{\dagger}(\bx)\phi(\bx) d\bx\ .
\end{align}
Here the pseudodifferential operator $\omega(i\nabla)$ acts as a Fourier multiplier through the formula
\begin{align}
    \omega(i\nabla)f(\bx) = \int_{\R^3} e^{i\bk\cdot\bx}\omega(\bk)\tilde{f}(\bk)\frac{d\bk}{(2\pi)^3}\ .
\end{align}

In order to obtain system \eqref{eq:dynamics1} from the Hamiltonian we must restrict ourself to the conserved subspace comprising vectors of the form
\begin{align}
    \vert\Psi\rangle = \int_{\R^3} \left(\psi(\bx,t)\phi^{\dagger}(\bx) + a(\bx,t)\rho(\bx)\sigma^{\dagger}(\bx) \vert 0\rangle\right)d^3 x\ .
\end{align}
Again, most terms are the same as in ~\cite{kraisler22} Appendix B. The one term which is different is
\begin{align}
    \langle 0\vert \phi(\bx) \int_{\R^3} \hbar\omega(i\nabla)\phi^{\dagger}(\by)\psi(\by,t)d^3 y \vert 0\rangle = \hbar\omega(i\nabla)\psi(\bx,t)\ .
\end{align}


\section{Derivation of the Kinetic Equations (\ref{eq:Kinetic1})}\label{appx:b}

In this appendix we derive the \Cref{eq:Kinetic1} satisfied by the functions $a_{\pm}$, which are the diagonal entries of $W_0$ in the basis $\{\bb_1(\bk),\bb_2(\bk)\}$. Having already addressed the equation at $O(1)$, we start with the remaining equations in the hierarchy \eqref{eq:osqrteps} and \eqref{eq:oeps}
\begin{multline}
    O(\sqrt{\eps}): \Tilde{A}_\eps (-(\bk - \bQ/2)/\eps) \Tilde{W}_1(\bx, \bQ, \bk, t) - \Tilde{W}_1(\bx, \bQ, \bk, t) \Tilde{A}_\eps (-(\bk + \bQ/2)/\eps) \\= 
    g \sqrt{\rho_0} \Tilde{\eta}(\bq) [W_0(\bx, \bk - \bQ/2, t)K^T - K W_0(\bx, \bk + \bQ/2, t)].
\end{multline}
\begin{multline}
   O(\eps): i\partial_tW_0 = LW_2 + g\sqrt{\rho_0} \int \frac{d^3 q}{(2\pi)^3} e^{i\bq\cdot\bX} \Tilde{\eta}(\bq) [K W_1(\bx, \bx, \bk + \bq/2, t) - W_1(\bx, \bx, \bk - \bq/2, t)K^T] \\
    + \int \frac{d^3 q}{(2\pi)^3} \frac{d^3 Q}{(2\pi)^3} e^{i\bq\cdot\bx + i\bQ\cdot\bX} [\Tilde{M}(\bq/2, -\bk + \bQ / 2) \tilde{W}_0(\bq, \bk, t) + \Tilde{W}_0(\bq, \bk, t) \Tilde{M}(-\bq/2, -\bk - \bQ / 2)],
\end{multline}
where 
\begin{align}
    L\We(\bx, \bk) &= \int \dq e^{i\bq \cdot \bx/\eps} \tilde{\eta}(\bq) [\We(\bx, \bk + \bq/2) - \We(\bx, \bk - \bq/2)] \ , \\
    \Tilde{M}(\bq/2, -\bk + \bQ / 2) &= \begin{bmatrix}
        \nabla \omega(-\bk + \bQ/2)\cdot \bq/2& 0 \\
        0 & 0
    \end{bmatrix}
\end{align}
Following a similar approach to \cite{kraisler22}, at $O(\sqrt{\eps})$, we can decompose $\Tilde{W}_1$ as:
\begin{align}
    \Tilde{W}_1 = \sum_{m, n}w_{mn}(\bx, \bQ, \bk, t) \bb_m(\bk - \bQ/2) \bb_n^T(\bk + \bQ/2)
\end{align}
\begin{multline}
    w_{mn}(\bx, \bQ, \bk, t) =\\= \frac{g^2 \rho_0 \Tilde{\eta}(\bQ)((\lambda_n(\bk + \bQ/2) - \Omega)a_m(\bx, \bp_1, t) - (\lambda_m(\bp_2) - \Omega)a_n(\bx, \bp_2, t))}{\sqrt{(\lambda_m(\bp_1) - \Omega)^2 + g^2 \rho_0} \sqrt{(\lambda_n(\bp_2) - \Omega)^2 + g^2 \rho_0} (\lambda_m(\bp_1) - \lambda_n(\bp_2) + i\theta)}.
\end{multline}
We focus on the $O(\eps)$ case:
\begin{multline}
    i\partial_tW_0 = LW_2 + g\sqrt{\rho_0} \int \frac{d^3 q}{(2\pi)^3} e^{i\bq\cdot\bX} \Tilde{\eta}(\bq) [K W_1(\bx, \bx, \bk + \bq/2, t) - W_1(\bx, \bx, \bk - \bq/2, t)K^T] \\
    + \int \frac{d^3 q}{(2\pi)^3} \frac{d^3 Q}{(2\pi)^3} e^{i\bq\cdot\bx + i\bQ\cdot\bX} [\Tilde{M}(\bq/2, -\bk + \bQ / 2) \tilde{W}_0(\bq, \bk, t) + \Tilde{W}_0(\bq, \bk, t) \Tilde{M}(-\bq/2, -\bk - \bQ / 2)]\ .
\end{multline}
Since $W_0$ has no $X$ term, we can suppose $Q = 0$, and transform $\Tilde{M}$. And so, we get:
\begin{align}
    \Tilde{M}(\bq/2, -\bk + \bQ / 2) - \Tilde{M}(-\bq/2, -\bk - \bQ / 2) = \nabla \omega(-\bk) \cdot \bq.
\end{align}
From
\begin{align}
\int \frac{d^3 q}{(2 \pi)^3} e^{i \bq \cdot \bx} \nabla \omega(-\bk) \cdot \bq f(\bq) = -i \nabla \omega(-\bk) \cdot \nabla_\bx f,
\end{align}
We get the equation:
\begin{multline}
    i \partial_t W_0(\bx, \bk, t) = LW_2 + M(\bx, \bk) W_0(\bx, \bk, t) + g \sqrt{\rho_0} \int \frac{d^3q}{(2\pi)^3} e^{i \bq \cdot \bX } \Tilde{\eta} (\bq) [KW_1(\bx, \bX, \bk + \bq/2, t) - \\
    W_1(\bx, \bX, \bk - \bq/2, t) K^T],
\end{multline}
\begin{align}
M(\bx, \bk) = \begin{bmatrix} -c i \nabla \omega(-\bk) \cdot \nabla_\bx & 0 \\ 0 & 0 \\ \end{bmatrix}.
\end{align}

We then multiply the left by $\bb_\pm^T(\bk)$ and the right by $\bb_\pm(\bk)$, and take the average with $\langle \bb_\pm^T LW_2 \bb_\pm \rangle = 0$. 

Then, since we know:
\begin{align}
    \langle \tilde{\eta}(\bq) \tilde{\eta}(\bQ)\rangle = (2\pi)^3 \tilde{C}(\bq) \delta(\bq + \bQ),
\end{align}
and
\begin{align}
    \lim_{\theta \rightarrow 0} \left( \frac{1}{x - i\theta} - \frac{1}{x + i\theta} \right) = 2\pi i \delta(x),
\end{align}
we get
\begin{multline}
    \partial_t a_+(\bx, \bk, t) = - c f_\pm(\bk) \nabla \omega(-\bk) \cdot \nabla_\bx a_\pm(\bx, \bk, t) + \\
    \langle \bb_+^T g\sqrt{\rho_0} \int \frac{d^3q}{(2\pi)^3} e^{i \bq \cdot \bX } \Tilde{\eta} (\bq) [KW_1(\bx, \bX, \bk + \bq/2, t) - W_1(\bx, \bX, \bk - \bq/2, t) K^T] \bb_+ \rangle.
\end{multline}
From which we can use the identity 
\begin{multline}
    \langle \bb_+^T g\sqrt{\rho_0} \int \frac{d^3q}{(2\pi)^3} e^{i \bq \cdot \bX } \Tilde{\eta} (\bq) [KW_1(\bx, \bX, \bk + \bq/2, t) - W_1(\bx, \bX, \bk - \bq/2, t) K^T] \bb_+ \rangle \\= \frac{2\pi (g^2 \rho_0 )^2(\lambda_+(\bk) - \Omega)^2}{((\lambda_+(\bq) - \Omega)^2 + g^2\rho_0)^2} \int \frac{d^3q}{(2\pi)^3} \Tilde{C}(\bk - \bq) \delta(\lambda_+(\bq) - \lambda_+(\bk))[a_+(\bx, \bk, t) - a_+(\bx, \bq, t)].
\end{multline}
Combining $(35)$ and $(37)$,
\begin{multline}
     \partial_t a_+(\bx, \bk, t) +  c f_\pm(\bk) \nabla \omega(-\bk) \cdot \nabla_\bx a_\pm(\bx, \bk, t) \\= \frac{2\pi (g^2 \rho_0 )^2(\lambda_+(\bk) - \Omega)^2}{((\lambda_+(\bq) - \Omega)^2 + g^2\rho_0)^2} \int \frac{d^3q}{(2\pi)^3} \Tilde{C}(\bk - \bq) \delta(\lambda_+(\bq) - \lambda_+(\bk))[a_+(\bx, \bk, t) - a_+(\bx, \bq, t)].
\end{multline}
Using the identity
\begin{align}
    \delta(\lambda_+ (\bq) - \lambda_+ (\bk)) = \frac{\delta(\omega(-\bq) - \omega(-\bk)) \sqrt{(c \omega(\bk) - \Omega)^2 + 4 g \rho_0}}{c(\lambda_+(\bk) - \Omega)},
\end{align}
\begin{multline}
     \partial_t a_\pm (\bx, \bk, t) +  c f_\pm (\bk) \nabla \omega(-\bk) \cdot \nabla_\bx a_\pm(\bx, \bk, t) = \\
     \mu_\pm(\bk) \int \frac{d^3q}{(2\pi)^3}\delta(\omega(-\bq) - \omega(-\bk)) \Tilde{C}(\bk - \bq) [a_\pm(\bx, \bk, t) - a_\pm(\bx, \bq, t)],
\end{multline}
we find
\begin{align}
    f_\pm(\bk) &= \frac{(\lambda_\pm(\bk) - \Omega)^2}{(\lambda_{\pm}(\bk) - \Omega)^2 + g^2\rho_0}, \\
    \mu_\pm(\bk) &= \frac{2\pi (g^2 \rho_0 )^2|\lambda_\pm (\bk) - \Omega|}{c((\lambda_\pm (\bq) - \Omega)^2 + g^2\rho_0)^2} \sqrt{(c \omega(-\bk) - \Omega)^2 + 4 g^2 \rho_0},
\end{align}
as desired.
\printbibliography
\end{document}